\documentclass[twocolumn,epsf,prb,amsmath,amssymb,showpacs,superscriptaddress]{revtex4}
\usepackage{graphicx}
\usepackage{amsmath}
\usepackage{amssymb}
\usepackage[T1]{fontenc}
\usepackage[usenames]{color}
\begin{document}
\title{Simplified model for the energy levels of quantum rings in single layer and bilayer graphene}

\author{M. Zarenia}
\affiliation{Departement Fysica, Universiteit Antwerpen, \\
Groenenborgerlaan 171, B-2020 Antwerpen, Belgium}
\author{J. Milton Pereira}
\affiliation{Departamento de F\'{\i}sica, Universidade
Federal do Cear\'a, Fortaleza, Cear\'a, $60455$-$760$, Brazil}
\author{A. Chaves}
\affiliation{Departamento de F\'{\i}sica, Universidade
Federal do Cear\'a, Fortaleza, Cear\'a, $60455$-$760$, Brazil}
\author{F.~M.~Peeters}\email{francois.peeters@ua.ac.be}
\affiliation{Departement Fysica, Universiteit Antwerpen, \\
Groenenborgerlaan 171, B-2020 Antwerpen, Belgium}
\affiliation{Departamento de F\'{\i}sica, Universidade
Federal do Cear\'a, Fortaleza, Cear\'a, $60455$-$760$, Brazil}
\author{G. A. Farias}
\affiliation{Departamento de F\'{\i}sica, Universidade
Federal do Cear\'a, Fortaleza, Cear\'a, $60455$-$760$, Brazil}

\begin{abstract}
Within a minimal model, we present analytical expressions for the eigenstates and eigenvalues of
carriers confined in quantum rings in monolayer and bilayer graphene. The calculations were performed
in the context of the continuum model, by solving the Dirac equation for a zero width ring geometry, i.e.
by freezing out the carrier radial motion. We include the effect of an external magnetic field and show the
appearance of Aharonov-Bohm oscillations and of a non-zero gap in the spectrum. Our minimal model gives insight
in the energy spectrum of graphene-based quantum rings and models different aspects of
finite width rings.
\end{abstract}
 \maketitle

\section{Introduction}
The investigation of low-dimensional solid state devices has
allowed the direct observation of quantum behavior in electron systems.
These effects arise due to the
confinement of carriers in structures that constrain
their movement along one or more directions, such as quantum wells, quantum wires and
quantum dots. One important class of such low-dimensional systems
are quantum rings, in which a particular type of confinement
together with phase coherence of the electron wavefunction
allows the observation of effects such as the Aharonov-Bohm \cite{AB} and Aharonov-Casher \cite{AC} effects.
Quantum rings have been extensively studied in semiconductor
systems, both experimentally and theoretically \cite{Bichier} and are expected to find application
in microelectronics, as well as in future quantum information devices.

In this paper we present analytical results for the eigenstates and
energy levels of ideal quantum rings created with graphene and
bilayer graphene. Graphene is an atomic layer of crystal carbon
which has been the target of intense scrutiny since it has been
experimentally produced \cite{novo3,novo4,shara,zhang}. Part of this
interest stems from the unusual properties of carriers in graphene,
caused by the gapless and approximately linear carrier spectrum,
together with possible technological applications, such as
transistors, gas sensors and transparent conducting materials in
e.g. photovoltaics. Additionally, it has been found that two coupled
graphene sheets, also known as bilayer graphene displays properties
that are distinct from single layer graphene as well as from
graphite. The carrier spectrum of bilayer graphene is gapless and
approximately parabolic at the vicinity of two points in the
Brillouin zone \cite{Bart,Ohta}. In particular, the spectrum is
strongly influenced by an external electric field perpendicular to
the bilayer, leading to the appearance of a gap \cite{McCann}. The
high quality of the single layer and bilayer graphene samples that
have been obtained, together with the large mean free path of
carriers suggests that phase coherence effects may be observable in
these systems. Recently, graphene-based quantum rings produced by
lithographic techniques have been investigated on single-layer
graphene \cite{Russo,Molitor}. These systems have been studied
theoretically by means of a tight-binding model, which does not
provide straightforward analytical solutions for the eigenstates and
eigenvalues \cite{Recher,Wurm,Luo,Rycerz,Bahamon}.
For bilayer graphene also
it was recently shown \cite{Zarenia} that is possible to electrostatically confine quantum rings with a finite width. The energy spectrum was obtained by solving the Dirac equation numerically.

In this paper we present a toy model that allows for analytical expressions for the energy levels
of quantum rings in single layer and bilayer graphene.
This model permits the description of several aspects of
the physics of graphene quantum rings without the additional complications of edge effects and finite
width of the quantum ring. We are able to obtain analytical expressions for the
energy spectrum and the corresponding wavefunctions, the persistent current, and the orbital momentum
as function of ring radius, total momentum and magnetic field, which can
be related to the numerical results obtained by other methods.

The paper is organized as follows: in section II
we present the theoretical model and numerical results for quantum rings in single layer graphene.
Similar results for bilayer graphene are given in
Sect. III. Section IV contains a summary of the main results and
the conclusions.

\section{Single layer graphene}

\subsection{Model}
The dynamics of carriers in the honeycomb lattice of covalent-bond
carbon atoms of single layer graphene can be described by the Dirac
Hamiltonian (valid for $E < 0.8$ eV). In the presence of a uniform
magnetic field ${\mathbf B} = B_{0} {\mathbf e}_z$ perpendicular to
the plane and finite mass term $\Delta$, which might be caused by an
interaction with the underlaying substrate \cite{McCann,Recher2} The
Hamiltonian in the valley isotropic form is given by \cite{Recher}:
\begin{equation}\label{eq1}
H=v_{F}(\mathbf{p}+e\mathbf{A}).\boldsymbol{\sigma}+ \tau\Delta \sigma_{z}
\end{equation}
where $\tau=+1$ corresponds to the $K$ point and $\tau=-1$ to the $K'$ point. ${\mathbf p}$ is the in-plane momentum
operator, $\mathbf A$ is the vector potential and $v_F \approx 1.0 \times 10^6$ m$/$s is the Fermi velocity, and $\boldsymbol{\sigma}= (\sigma_x,\sigma_y,\sigma_z)$ is the pseudospin operator with components given by Pauli matrices.
The eigenstates of Eq. (\ref{eq1}) are two-component spinors which, in polar coordinates is given by
\begin{equation}\label{eq2}
 \Psi(\rho,\phi)=\left(
        \begin{array}{c}
        \phi_{A}(\rho)e^{im\phi} \\
        i\phi_{B}(\rho)e^{i(m+1)\phi} \\
            \end{array}
      \right).
\end{equation}
where $m$ is the angular momentum label.
We follow the earlier very successful approach \cite{Meijer,Molnar} of ideal one-dimensional (1D) quantum rings in semiconductors with
spin-orbit interaction where the Schr\"odinger equation was simplified by discarding the radial variation
of the electron wave function.
Thus, in the case of an ideal ring with radius $R$, the momentum
of the carriers in the radial direction is zero. We treat the radial parts of the spinors in Eq. (\ref{eq2})
as a constant
\begin{equation}\label{eq3}
 \Psi(R,\phi)=\left(
        \begin{array}{c}
        \phi_{A}(R)e^{im\phi} \\
        i\phi_{B}(R)e^{i(m+1)\phi} \\
            \end{array}
      \right).
\end{equation}

Because the radial motion is frozen in our model there will be no radial current and the persistent current will be purely in the angular direction. By
solving $H\Psi(R,\phi)=E\Psi(R,\phi)$ and using the symmetric gauge ${\bf A}=(0,B_{0}\rho/2,0)$,
we obtain
\begin{eqnarray}\label{eq4}
&&(m+1+\beta)\phi_{B}(R)=(\epsilon-\tau\delta)\phi_{A}(R),\cr &&\cr
&&(m+\beta)\phi_{A}(R)=(\epsilon+\tau\delta)\phi_{B}(R),
\end{eqnarray}
where the energy and mass terms are written in dimensionless units as $\epsilon=E/E_{0}$, $\delta=\Delta/E_{0}$ with $E_{0}=\hbar v_{F}/R$.
The parameter $\beta=(eB_{0}/2\hbar)R^{2}$ can be expressed as $\beta=\Phi/\Phi_{0}$ with $\Phi = \pi R^2 B_{0}$ being the magnetic flux threading the ring and $\Phi_{0}=h/e$
the quantum of magnetic flux.
The homogeneous set of equations (\ref{eq4}) is solvable for the energies
\begin{equation}\label{eq5}
\epsilon=\pm\sqrt{(m+\beta+1)(m+\beta)+\delta^{2}}.
\end{equation}
%
This energy can also be written as
\begin{equation}
\epsilon=\pm\sqrt{(m-m_-)(m-m_+)}
\end{equation}
where
\begin{equation}\label{eq51}
\displaystyle{m_{\pm}=-(\beta+\frac{1}{2})\pm\sqrt{\frac{1}{4}-\delta^{2}}}.
\end{equation}
Note that the energy spectrum for an ideal single layer quantum ring for both $K$ and $K'$ points are the same. For $|\delta|> 1/2$ we have that $m_+ = m_-^*$ is complex and $\epsilon$ is real for any value of $\beta$.
In the region $-\frac{1}{2}<\delta<\frac{1}{2}$ the energy is real, except for $m_{-}<m<m_{+}$, when
the energy is complex. For the gapless case, i.e. $\delta = 0$, we have $m_+ = -\beta$ and $m_- = -\beta-1$
and the energy is real when $m < -\beta - 1$ or when $m > -\beta$ and imaginary otherwise.

The wavefunctions are eigenfunctions of the total angular momentum operator given by the sum
of orbital angular momentum $L_z$ and a term describing the pseudo-spin $S_z$
\begin{equation}
J_z = L_z + \hbar S_z,
\end{equation}
where $S_z = (1/2) \sigma_z$, with $\sigma_z$ being one of the Pauli matrices and the eigenvalues of $J_{z}$ operator become $[m+(1/2)]\hbar$.

The current is obtained using $j_{x,y} = v_F[\psi^\dagger \sigma_{x,y}\psi]$. The total angular current $j=v_{F}[\psi^\dagger \sigma_{\phi}\psi]$ can be calculated using the fact that $\sigma_\phi = \xi(\phi) \sigma_y$, where
\begin{equation}\label{eqq}
\xi(\phi) =
\begin{pmatrix}
  e^{-i\phi} & 0  \\
  0 & e^{i\phi}
\end{pmatrix}
.
\end{equation}
The current for the electrons in the K-valley becomes
\begin{equation}\label{eqq1}
j_{K} =v_F(\phi_A^*\phi_B + \phi_B^*\phi_A).
\end{equation}
The total current is then given by $j=j_{K}+j_{K'}$. The radial part of the two spinor components are
\begin{equation}\label{eqq2}
\phi_A(R) = 1, \quad \phi_B(R)= \frac{m+\beta}{\epsilon + \tau\delta}.
\end{equation}
Notice that the radial current can be calculated using $j_{r} = v_F[\psi^\dagger \xi(\phi) \sigma_x\psi]$ which leads to the the following relation,
\begin{equation}\label{eq}
j_{r} =iv_F(\phi_A^*\phi_B - \phi_B^*\phi_A),
\end{equation}
where, in the case of our ideal ring we have $j_{r}=0$. From Eqs. (\ref{eqq1}) and (\ref{eqq2}), one can find the following expression for the total angular current of a single layer quantum ring
\begin{equation}\label{j}
j=\frac{4v_{F}\epsilon(m+\beta)}{\epsilon^2 - \delta^2}.
\end{equation}
One can rewrite Eq. (\ref{j}) in the following form
\begin{equation}\label{eq}
\frac{j}{v_{F}}=\big(\frac{\partial \epsilon}{\partial \beta}\big)_{K} + \big(\frac{\partial \epsilon}{\partial \beta}\big)_{K'} +
\frac{2(m+\beta)(\epsilon^2+\delta^2)-(\epsilon^2-\delta^2)}{\epsilon(\epsilon^2-\delta^2)}.
\end{equation}
Since for the ground state energy $\sqrt{\delta^2-1/4}\leq\epsilon\leq\delta$ and $-1/2\leq m+\beta\leq 0$ the last term in Eq. (13) is much smaller than the derivatives of the energy with respect to the flux ($\Phi$) and oscillates around zero. Note that in 1D semiconductor rings the current is exactly given by $\partial E/\partial \Phi$, which is thus different from graphene where we have approximately
\begin{equation}\label{eq}
\frac{j}{v_{F}}\simeq\big(\frac{\partial \epsilon}{\partial \beta}\big)_{K} + \big(\frac{\partial \epsilon}{\partial \beta}\big)_{K'}.
\end{equation}
with $\beta=\Phi/\Phi_{0}$.

\subsection{Results}
\begin{figure}
\centering
\includegraphics[width=9.3 cm]{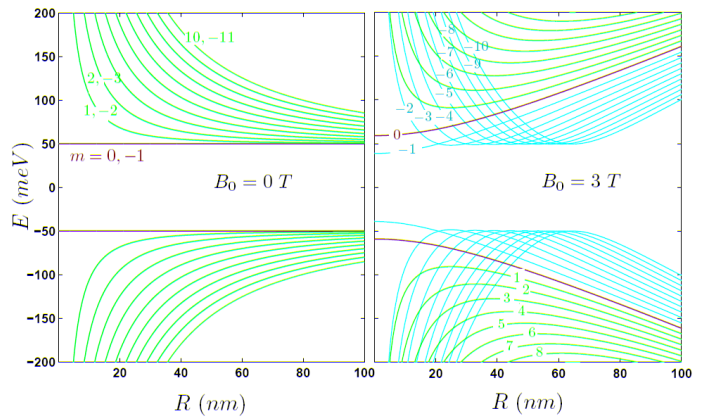}
\caption{(Color online) Energy levels with $m=-10,...,10$ of single layer graphene quantum ring as function of ring radius $R$ for $B_{0}=0$ T (left panel) and $B_{0}=3$ T (right panel) when the mass term is $\Delta=50$ meV.} \label{fig1}
\end{figure}

%
The energies as function of ring radius R are shown in Fig. \ref{fig1}, for $\Delta = 50$ meV, with $-10\leq m\leq-1$ (magenta curves),
$1\leq m\leq10$ (blue curves), and $m=0$ (green curves). In the absence of an external magnetic field, the energy is given by
$E=\pm\sqrt{m(m+1)(\hbar v_F/R)^2 + \Delta^2}$ and
the energy branches have a $1/R$ dependence and approach $E \rightarrow \pm\Delta$ for very large radii.
Note that for $m=0$ and $m=-1$ the energy $E = \pm \Delta$ is independent of $R$ and all branches are two-fold
degenerate.
For non-zero magnetic field ($B = 3$T), the right panel shows
that the branches have an approximately linear dependence on the ring radius for large $R$, in particular we have
$E\simeq \pm\sqrt{(\alpha R)^2 + \Delta^2}$, with $\alpha = v_F e B_0/2$. For small radii, $E \simeq \pm \hbar\sqrt{m(m+1)}/R$ and
all branches diverge as $1/R$, except for  $m=0$ and $m=-1$. In those cases when $R \rightarrow 0$ we have for $m = 0$ the result
$E=\pm\sqrt{\Delta^2 + \alpha \hbar v_F}$, while $m = -1$ gives $E=\pm\sqrt{\Delta^2 - \alpha \hbar v_F}$.

\begin{figure}
\centering
\includegraphics[width=8 cm]{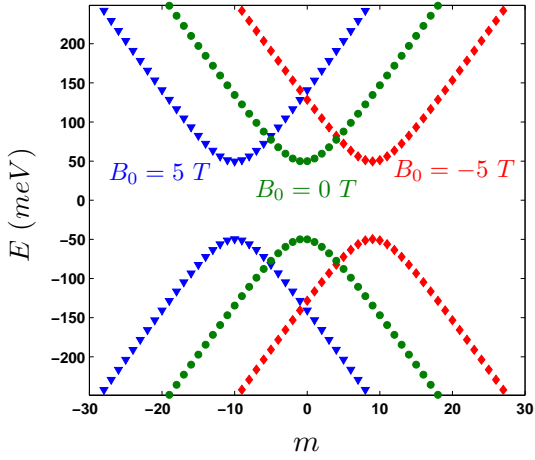}
\caption{(Color online) Energy levels of a single layer graphene quantum ring as function of the quantum number $m$ for $B_{0}=-5,0,5$ T with $\Delta=50$ meV and $R=50$ nm.} \label{fig2}
\end{figure}

Figure \ref{fig2} presents results for the energy as function of total angular momentum index $m$,
for $\Delta = 50$ meV, $R = 50$ nm and for three different values of magnetic field, namely $B_0 = -5$ T (diamonds), $B_0 = 0$ T (circles) and
$B_0 = 5$ T (triangles). Notice that for a given $B_0$ the electron energy obtains a minimum for a particular $m$, i.e. for
$B_0 = 0$ ($5$ T, $-5$ T) it is $m = 0$ ($9$, $-10$). In fact it is given by $m = -(\Phi/\Phi_0 + 1/2)$ and is independent
of $\Delta$.

\begin{figure}
\centering
\includegraphics[width=8.75 cm]{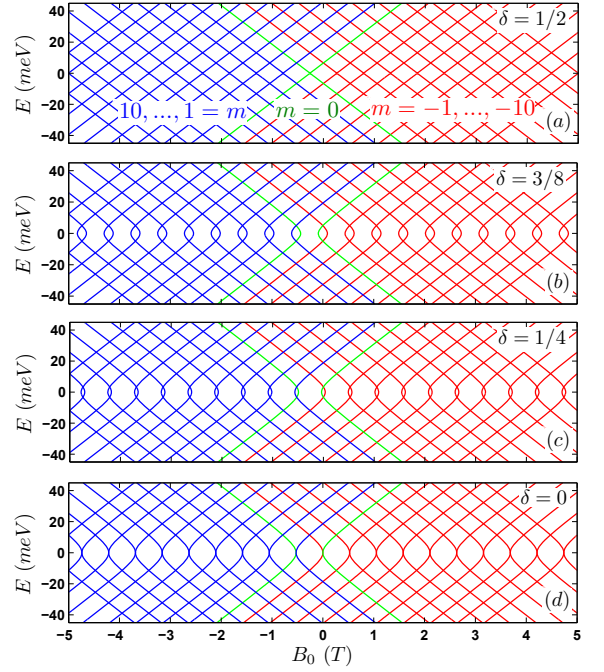}
\caption{(Color online) Electron and hole energy levels of a single layer graphene quantum ring as function of external magnetic field $B_{0}$ for (a) $\delta=1/2$, (b) $\delta=3/8$, (c) $\delta=1/4$ and (d) $\delta=0$ with $R=50$ nm, and total angular quantum number $-10\leq m\leq-1$ (red curves), $1\leq m\leq10$ (blue curves) and $m=0$ (green curves).}
\label{fig3}
\end{figure}

\begin{figure}
\centering
\includegraphics[width=8 cm]{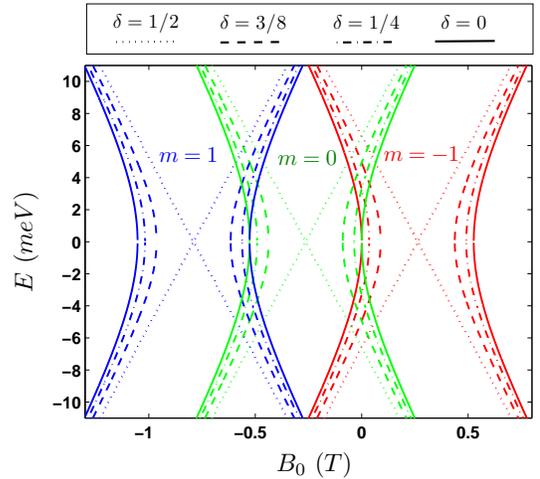}
\caption{(Color online) The same as Fig. 3, for $m=-1$ (red curves), $m=1$ (blue curves) and $m=0$ (green curves) and different values of dimensionless mass term $\delta$.}
\label{fig4}
\end{figure}

The energy levels as function of the external magnetic field are shown in Fig. \ref{fig3}, for a quantum ring with (a) $\delta=1/2$, (b) $\delta=3/8$, (c) $\delta=1/4$ and (d) $\delta=0$ with $R=50~nm$ for $-10\leq m\leq-1$ (red curves), $1\leq m\leq10$ (blue curves), and $m=0$ (green curves).
The magnetic field dependence of the spectrum becomes evident if one rewrites Eq. (5) as $\epsilon^2 - [(m+\Phi/\Phi_0)+1/2]^2 = \delta^2-1/4$.
Thus, for the special case of $\delta =\Delta/E_0= 1/2$ the gap is zero and the energy levels are straight lines given by $\epsilon = \pm(m+1/2+\Phi/\Phi_0)$. The energy spectra for $\delta > 1/2$ resemble those found earlier by Recher {\it et al.} \cite{Recher} in the case of a finite width graphene
ring with infinite mass boundary conditions. An enlargement of Fig. \ref{fig3} around $E=0$ is shown in Fig. \ref{fig4}. The spectrum has an interesting magnetic field dependence with decreasing $\delta$. 
For $\delta =0$ the double degeneracy is restored at $E = 0$. This behavior can be easily illustrated by considering $m = 0$. The energy in this case is $\epsilon = \pm \sqrt{\beta(\beta + 1)+ \delta^2}$ which for $\delta = 1/2$ becomes $\epsilon = \pm (\beta + 1/2)$ while for $\delta = 0$ it
is $\epsilon = \pm \sqrt{\beta(\beta +1)}$ and thus $\epsilon \simeq \pm \sqrt\beta$ for $\beta \simeq 0$.

\begin{figure}
\centering
\includegraphics[width=8 cm]{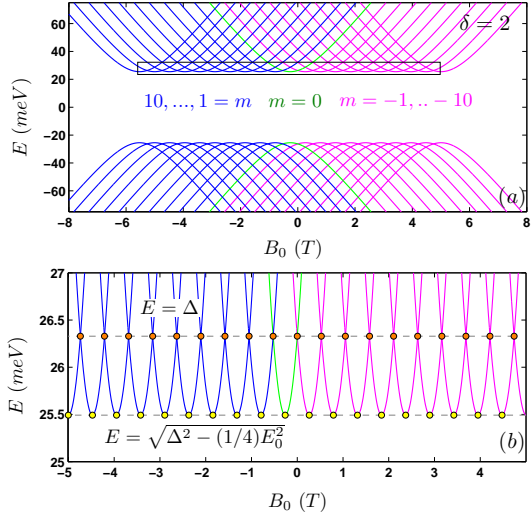}
\caption{(Color online) (a) Electron and hole energy levels of a single layer graphene quantum ring as function of external magnetic field $B_{0}$ for $\delta=2$ and $R=50$ nm. (b) An Enlargement of the region which is shown in (a) by a rectangle.}\label{fig5}
\end{figure}

\begin{figure}
\centering
\includegraphics[width=9.3 cm]{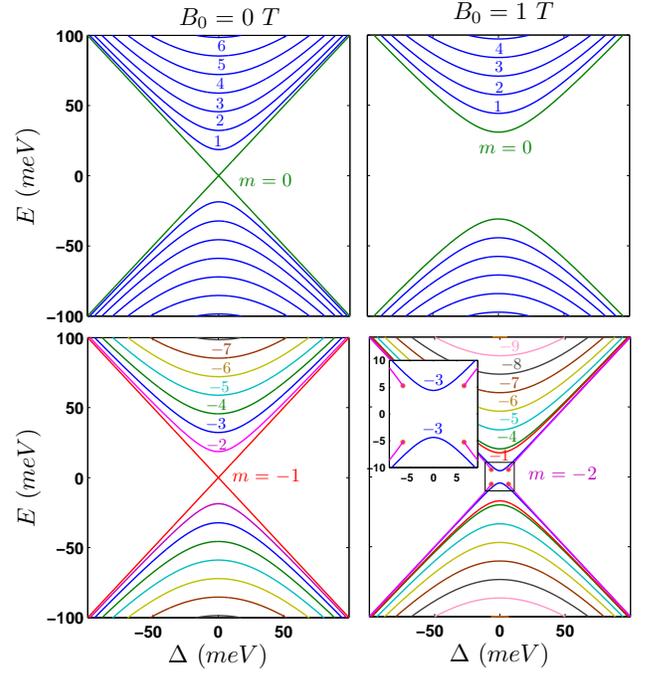}
\caption{(Color online) Lowest energy levels of a single layer graphene quantum ring as function of the mass term $\Delta$ with $B_{0}=0$ T (left panels) and $B_{0}=1$ T (right panels) for $m\geq0$ (upper panels) and $m<0$ (lower panels) with $R=50$ nm.} \label{fig6}
\end{figure}

In Fig. \ref{fig5}(a) the energy spectrum is plotted vs magnetic field for $\delta = 2$. Where, the energy has a hyperbolic dependence on the applied magnetic field with minima at $\Phi/\Phi_0 = -m-1/2$ and a gap of $\Delta \epsilon = 2\sqrt{\delta^2-1/4}$. The exact location of the transitions (orange dots) and the location of the minima points (yellow dots) in the energy spectrum is clarified in Fig. \ref{fig5}(b).
The dependence of the energy levels on the gap parameter $\Delta$ is shown in Fig. \ref{fig6}, for zero magnetic field (left panels) and $B_0 = 1$ T (right panels). When $m\geq0$ (upper panels) and $m<0$ (lower panels). For $B_0=0$ T the energy levels are two fold degenerate where $E(m)=-E(-m-1)$. When a magnetic field is applied an energy gap is opened (see right-bottom panel in Fig. 6). Notice also that the $m=-2$ level only exists for $\Delta\geq E_{0}/2$, i. e. for $\Delta<E_{0}/2$ there is no real energy solution when $m=-2$.

\begin{figure}
\centering
\includegraphics[width=8 cm]{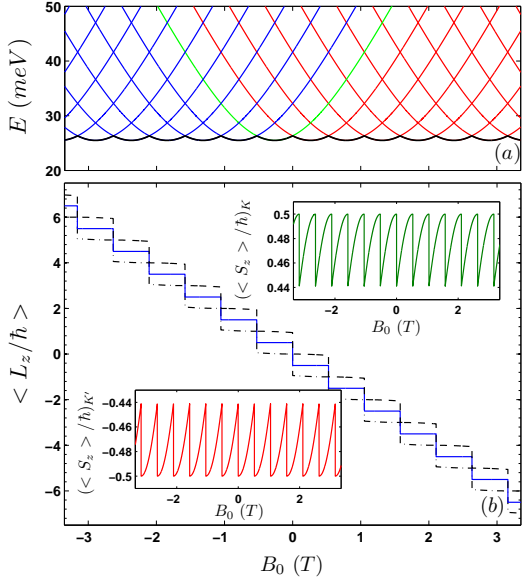}
\caption{(Color online) (a) Electron energy levels of a graphene single layer quantum ring as function of external magnetic field $B_{0}$ for the same parameters as used in Fig. 5. Black curve shows the ground state energy (b) Ground state expectation value of $L_{z}/\hbar$ as function of magnetic field for both $K$ (black dashed curve) and $K'$ valley (black dash-dotted curve). Expectation value of $S_{z}/\hbar$ versus magnetic field is plotted in upper inset for $K$-valley and in lower inset for $K'$-valley. Blue solid curve shows the expectation value $<J_z>$ which is the same for both valleyes.} \label{fig7}
\end{figure}

The corresponding ground state expectation values for the operators in Eq. (8) are plotted as function of the magnetic field in Fig. \ref{fig7}(b) for both $K$ (black dashed curve) and $K'$ valley (black dash-dotted curve). Notice that for the $K$-valley $<L_{z}>\simeq m\hbar$ and $<S_z>\simeq \hbar/2$ whereas in the $K'$-valley $<L_{z}>\simeq (m+1)\hbar$ and $<S_z>\simeq -\hbar/2$. Thus for both the $K$ valley and the $K'$ valley $<J_z>\simeq [m+(1/2)]\hbar$ which is approximately quantized and on the average its value decreases linearly with the applied magnetic field.

\begin{figure}
\centering
\includegraphics[width=8 cm]{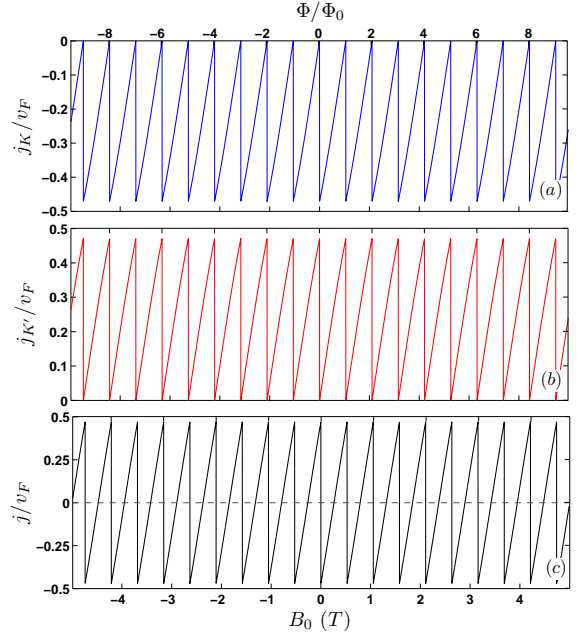}
\caption{The angular current density in the (a) $K$-valley, (b) $K'$-valley and (c) the total current density of a monolayer graphene quantum ring as function of external magnetic field $B_{0}$ for the ground state energy shown by the black curve in Fig. \ref{fig7}(a).} \label{fig8}
\end{figure}

The angular current density for a single layer graphene quantum ring is shown in Fig. \ref{fig8}(c). Note that the contribution from the $K$-valley $j_{K}$ (Fig. \ref{fig8}(a)) and the $K'$-valley $j_{K'}$ (Fig. \ref{fig8}(b)) are not the same, they have opposite sign and oscillate in phase around a nonzero average value $-\tau v_{F}/4$. The reason is that if for given energy we have electrons in the $K$-valley, the corresponding particles in the $K'$-valley will behave as holes.
The persistent current is a sawtooth shaped oscillating function of the magnetic field with period $\Delta B_{0}=\Phi_{0}/\pi R^{2}$. This behavior is quantitatively very similar to those found for the standard Aharanov-Bohm oscillations in metallic and semiconductor quantum rings.
%


\section{Bilayer graphene}
\subsection{Model}
In the case of bilayer graphene the Hamiltonian in the vicinity of the $K$ point, is given by\cite{McCann}
\begin{equation}\label{eq6}
H=\left(
  \begin{array}{cccc}
  \tau U_{1} &\pi& t & 0 \\
  \pi^{\dag}& \tau U_{1} & 0 & 0 \\
  t & 0 & \tau U_{2} &\pi^{\dag} \\
  0 & 0 & \pi& \tau U_{2} \\
  \end{array}
\right)
\end{equation}
where $\tau=\pm1$ distinguishes the two $K$ and $K'$ valleys. $t \simeq 400$ meV is the interlayer coupling term, $\pi =
v_F[(p_x+eA_x) + i(p_y+eA_y)]$, $U_1$ and $U_2$ are the potentials, respectively, at the two graphene layers. Here we do not include any mass term because the gate potential across the bilayer is able to open an energy gap in
the spectrum.
\cite{McCann} 
 The eigenstates of the Hamiltonian (\ref{eq6}), are four-component spinors $\Psi(r,\phi) =
[\phi_{A}(\rho)e^{im\phi} \, , \, i\phi_{B}(\rho)e^{i(m-1)\phi}\, , \, \phi_{C}(\rho)e^{im\phi}\, , \, i\phi_{D}(\rho)e^{i(m+1)\phi}]^T$
(see Ref. [\onlinecite{Milton}]). Following our earlier approach for an ideal ring with radius $R$, the wave function becomes:
\begin{equation}\label{eq7}
 \Psi=\left(
        \begin{array}{c}
        \phi_{A}(R)e^{im\phi} \\
          i\phi_{B}(R)e^{i(m-1)\phi} \\
           \phi_{C}(R)e^{im\phi} \\
           i\phi_{D}(R)e^{i(m+1)\phi} \\
        \end{array}
      \right).
\end{equation}
We use the symmetric gauge and obtain the following set of coupled algebraic equations
\begin{eqnarray}\label{eq8}
&&-(\epsilon-\tau u_{1}) \phi_{A}(R)- (m+\beta-1)\phi_{B}(R) + t'\phi_{C}(R)=0,\cr &&\cr
&&(m+\beta) \phi_{A}(R)+(\epsilon-\tau u_{1}) \phi_{B}(R) =0,\cr &&\cr &&
t' \phi_{A}(R)-(\epsilon-\tau u_{2}) \phi_{C}(R)+(m+\beta+1)\phi_{D}(R)=0,\cr &&\cr &&
(m+\beta) \phi_{C}(R)-(\epsilon-\tau u_{2})\phi_{D}(R) =0.
\end{eqnarray}
where, $t'=t/E_{0}$ and $u_{1,2}=U_{1,2}/E_{0}$ are in dimensionless units. After some straightforward algebra we obtain the following polynomial
equation that determines the energy spectrum
\begin{eqnarray}\label{eq9}
&&(\epsilon-\tau u_{1})^{2}\bigr[(\epsilon-\tau u_{2})^{2}-(m+\beta)(m+\beta+1)\bigl]\cr &&\cr
&&-(m+\beta)(m+\beta-1)\bigr[(\epsilon-\tau u_{2})^{2}-(m+\beta)(m+\beta+1)\bigl]\cr &&\cr
&&-(\epsilon-\tau u_{1})(\epsilon-\tau u_{2})t'^{2}=0.
\end{eqnarray}
\begin{figure}
\centering
\includegraphics[width=8 cm]{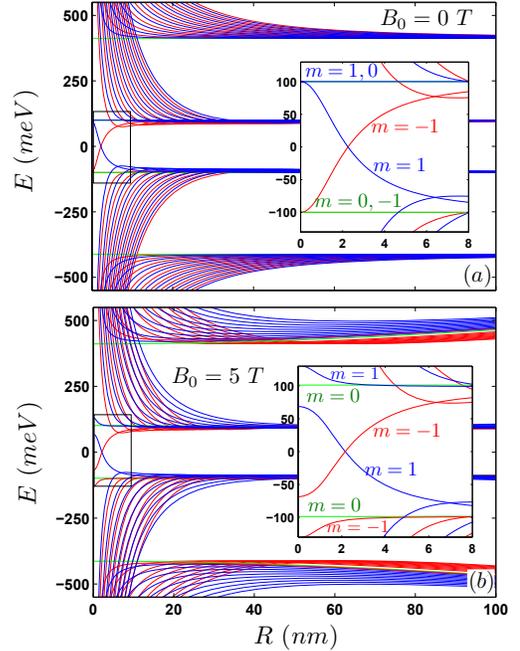}
\caption{(Color online) Lowest energy levels of a bilayer graphene quantum ring as function of ring radius $R$ with (a) $B_{0}=0$ T and (b) $B_{0}=5$ T for $U_{b}=100$ meV and total angular quantum number $-10\leq m\leq-1$ (red curves), $1\leq m\leq10$ (blue curves) and $m=0$ (green curves). The insets are an enlargement of the small energy and small $R$ region.}
\label{fig10}
\end{figure}
\begin{figure}
\centering
\includegraphics[width=8 cm]{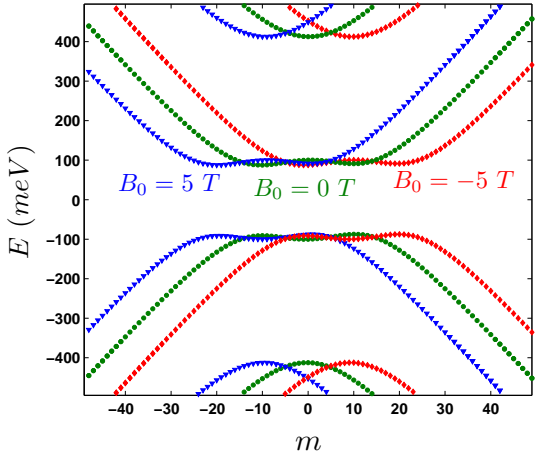}
\caption{(Color online) Lowest energy levels of a graphene bilayer quantum ring as function of total angular momentum label $m$ for $B_{0}=-5,0,5$ T with $U_{b}=100$ meV and $R=50$ nm.} \label{fig11}
\end{figure}
\begin{figure}
\centering
\includegraphics[width=8 cm]{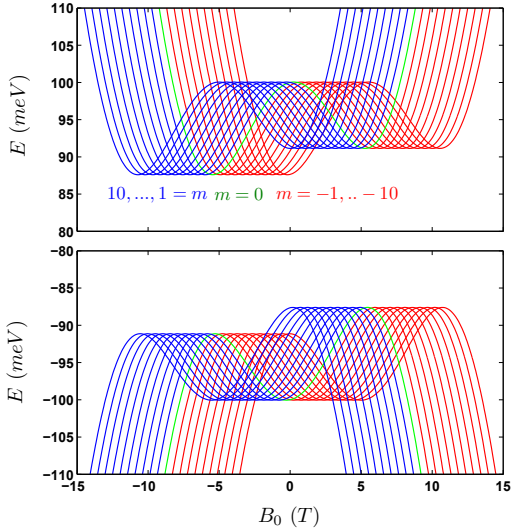}
\caption{(Color online) Electron and hole energy states of a graphene bilayer quantum ring as function of external magnetic field $B_{0}$ for $U_{b}=100$ meV and $R=50$ nm. The energy levels are shown for the quantum numbers $-10\leq m\leq-1$ (red curves), $1\leq m\leq10$ (blue curves) and $m=0$ (green curves).}
\label{fig12}
\end{figure}
After introducing the average potential $u = (u_1+u_2)/2$ and half the potential difference $\delta = (u_1-u_2)/2$ we can rewrite this quartic algebraic equation in
a more comprehensive form:
\begin{eqnarray}
&&s^4 - 2s^2[(m+\beta)^2 + \delta^2+(t')^2/2] \cr &&\cr
&&+4s \tau \delta (m+\beta)+ (m+\beta)^2
[(m+\beta)^2-1]\cr &&\cr
&&-2\delta^2[(m+\beta)^2-(t')^2/2]+\delta^4=0,
\end{eqnarray}
where $s=\epsilon - \tau u$ is the energy shifted by the average potential. In the next
section we report the results for the case of $U_{1}=-U_{2}=U_{b}$ where, the average potential $u$ is zero.
In the limit $\delta \rightarrow 0$, the quartic equation is reduced to
a quadratic equation in $s^2$ and we obtain the real solutions
\begin{eqnarray}
s_{\pm}^2&=&(m+\beta)^2+(t')^2/2\cr &&\cr
&&\pm \sqrt{(t')^4/4+(m+\beta)^2(1+t')^2},
\end{eqnarray}
which results in four solutions for the energy. These are real when $|m+\beta|\geq 1$.
In the opposite case of $|m+\beta|< 1$ (or equivalently $-1+\beta < m < 1 - \beta$)
we have $s_-^2 < 0$ and consequently the corresponding energies are imaginary.
In the limit of $t' >> m +\beta$ we obtain $s_-^2 = (m+\beta)^2[(m+\beta)^2-1]/(t')^2$
and thus the low energy solutions are given by
\begin{eqnarray}\label{eq11}
s\simeq \pm\frac{1}{t'}\sqrt{(m+\beta)^{2}((m+\beta)^{2}-1)}.
\end{eqnarray}

For bilayer graphene, the wavefunctions are eigenfunctions of the following operator:
\begin{equation}
J_z = L_z + \hbar \tau_z + \hbar S_z ,
\end{equation}
where now
\begin{equation}
\tau_z = \frac{1}{2}
\begin{pmatrix}
  -{\mathbf I} & 0  \\
  0 & {\mathbf I}
\end{pmatrix}
\quad , \quad
S_z = \frac{1}{2}
\begin{pmatrix}
  \sigma_z & 0  \\
  0 & -\sigma_z
\end{pmatrix}
,
\end{equation}
are $4\times 4$ matrices.

In bilayer graphene the components of the current density are given by
\begin{equation}
j_x =v_{F}\left[\psi^{\dag}
\begin{pmatrix}
  {\sigma_{x}} & 0  \\
  0 & {\sigma_{x}}
\end{pmatrix}\psi\right]
\quad , \quad
j_y = v_{F}\left[\psi^{\dag}
\begin{pmatrix}
  -\sigma_y & 0  \\
  0 & \sigma_y
\end{pmatrix}\psi\right].
\end{equation}
The angular current can be calculated from the following relation,
\begin{equation}
j=v_{F}\left[\psi^{\dag}\begin{pmatrix}
  \sigma_y^{\ast}\xi(\phi) & 0  \\
  0 & \xi(\phi)\sigma_y
\end{pmatrix}\psi\right].
\end{equation}
Where, $\xi(\phi)$ is given by Eq. (\ref{eqq}). We obtain for the angular current in the K-valley,
\begin{equation}\label{eqJB}
j_{K}=v_F(\phi_C^*\phi_D + \phi_D^*\phi_C-\phi_A^*\phi_B - \phi_B^*\phi_A),
\end{equation}
and the total angular current is given by $j=j_{K}+j_{K'}$. Where, the four spinor components are:
\begin{eqnarray}
&&\phi_A(R) = 1,\cr &&\cr &&\phi_B(R)= -\frac{m+\beta}{\epsilon - \tau u_{1}},\cr &&\cr
&&\phi_C(R) = \frac{(\epsilon - \tau u_{1})^{2}-(m+\beta)(m+\beta-1)}{t'(\epsilon - \tau u_{1})},\cr &&\cr
&&\phi_D(R)= \frac{(m+\beta)[(\epsilon - \tau u_{1})^{2}-(m+\beta)(m+\beta-1)]}{t'(\epsilon - \tau u_{1})(\epsilon - \tau u_{2})}.\cr &&\cr &&
\end{eqnarray}
Note that the radial current can be calculated through
\begin{equation}
j_{r}=v_{F}\left[\psi^{\dag}\begin{pmatrix}
  \sigma_x\xi(\phi) & 0  \\
  0 & \xi(\phi)\sigma_x
\end{pmatrix}\psi\right],
\end{equation}
where, $j_{r}=iv_F(\phi_A^*\phi_B - \phi_B^*\phi_A+\phi_C^*\phi_D - \phi_D^*\phi_C)=0$ for the case of an ideal ring. Using Eq. (\ref{eqJB}), the total current density becomes,
\begin{eqnarray}
&&j=\sum_{\tau=\pm1}\cr &&\cr&&\frac{2v_F(m+\beta)}{\epsilon-\tau u_{1}}\left[1+\frac{\big[(\epsilon - \tau u_{1})^{2}-(m+\beta)(m+\beta-1)\big]^{2}}{t'^{2}(\epsilon - \tau u_{1})(\epsilon - \tau u_{2})}\right].\cr &&
\end{eqnarray}

\subsection{Results}

The dependence of the spectrum on the ring radius, for $B_0 = 0$ T
(upper panel) and $B_0 = 5$ T (lower panel) is shown in Fig.
\ref{fig10}, for a gate potential $U_b = 100$ meV, which for $B = 0$
T opens up a gap in the energy spectrum. As compared to the single
layer quantum ring results of Fig. \ref{fig1}, we find two main
differences: i) for $R \rightarrow 0$ there are two states inside
the gap, and ii) we have a second set of levels that for large $R$
are displaced in energy by $t$. In the limit $R\rightarrow0$ the
most important term in the dispersion relation is
$(m+\beta)^{2}\bigr[(m+\beta)^{2}-1\bigl]$. For $m=-1,0,1$ the
behavior of the spectrum is different and the corresponding energy
levels don't diverge when $R\rightarrow0$. The same behavior was
found for the single layer results, but only for $m=0,-1$.
Previously, we found that for rings with finite width
\cite{Zarenia} the spectrum exhibits anti-crossing points which arise due to the overlap
of gate-confined and
magnetically-confined states. In the present model the carriers motion along the radial direction
is neglected and consequently we have level crossings instead of anti-crossing points
in the spectrum.
\begin{figure}
\centering
\includegraphics[width=8 cm]{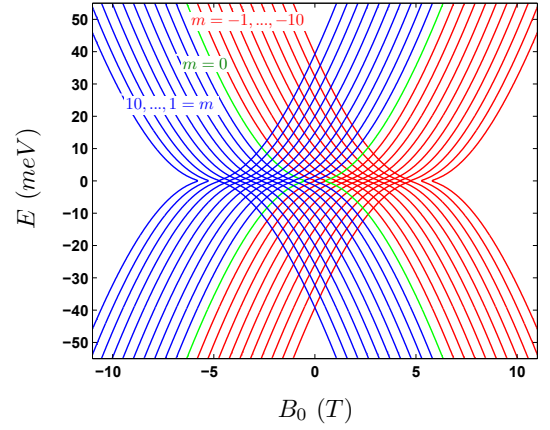}
\caption{ (Color online) The same as Fig. \ref{fig12}, but for $U_{b}=0$ meV.} \label{fig13}
\end{figure}
\begin{figure}
\centering
\includegraphics[width=9.3 cm]{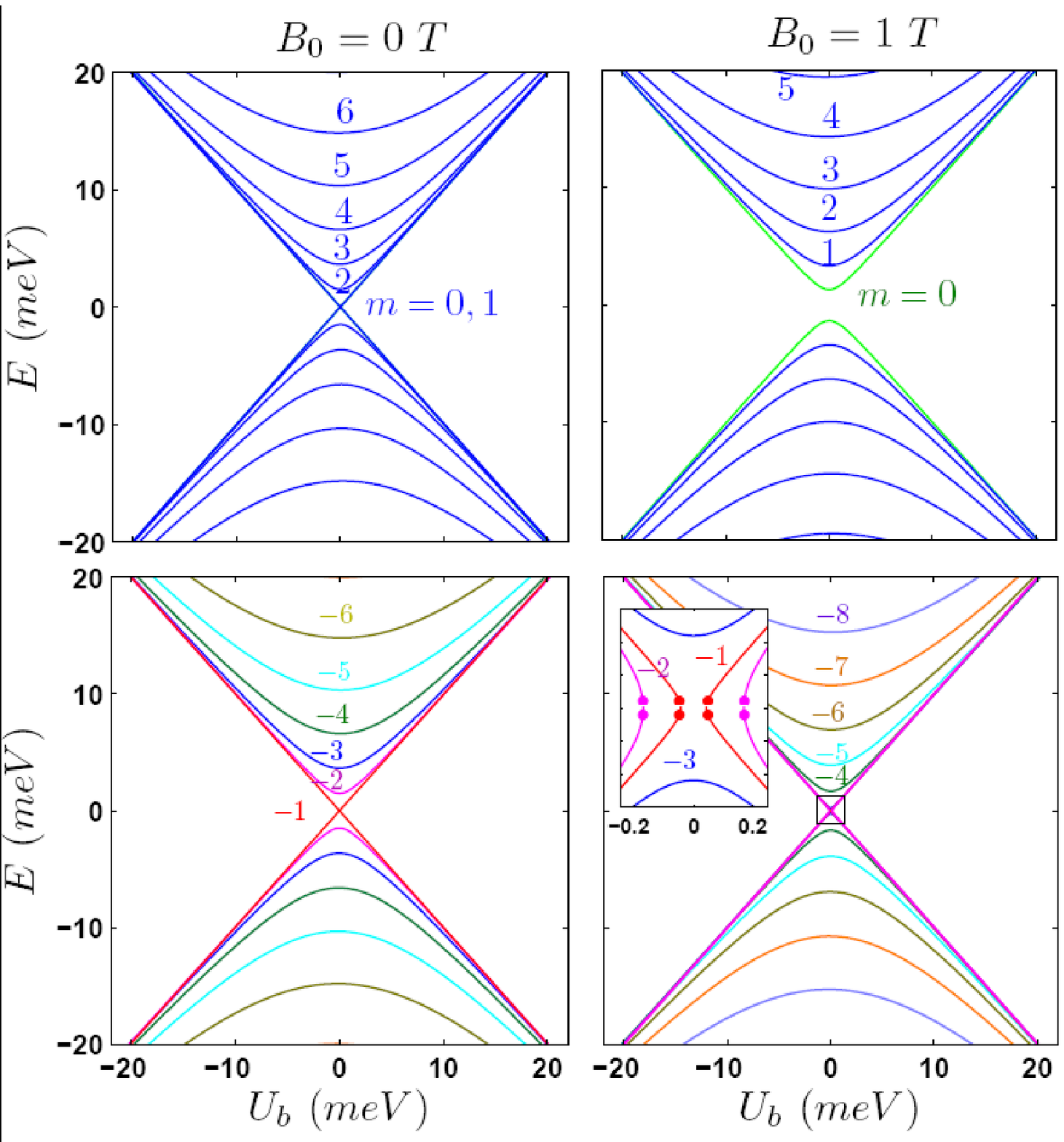}
\caption{(Color online) Lowest energy levels of a bilayer graphene quantum ring as function of the gate potential $U_{b}$ when $B_{0}=0$ T (left panels) and $B_{0}=1$ T (right panels) for $m\geq0$ (upper panels) and $m<0$ (Lower panels) with $R=50$ nm.} \label{fig14}
\end{figure}
The dependence of the energy eigenstates on the angular momentum index $m$ is displayed in Fig. \ref{fig11}
for $U_b = 100$ meV, $R = 50$ nm, with $B_0 = -5$ T (diamonds), $0$ T (circles)
and $5$ T (triangles). Due to the finite bias in this case, the fourth-order character of the
dispersion Eq. (\ref{eq9}) causes the curves to exhibit a Mexican hat shape. The energy minima for $B_{0}=-5, 0, 5$ T are
respectively given by $m=-1, -10, -20$.
In Fig. \ref{fig12} the energy levels are plotted as function of magnetic field, for a quantum ring
with $U_b = 100$ meV, $R=50~nm$, and for $-10\leq m\leq-1$ (red curves), $1\leq m\leq10$ (blue curves),
and $m=0$ (green curves). These results are very similar to those found for a finite width ring and
exhibit two local minima that are separated by a saddle point. In the case of finite width quantum rings
there are additional energy levels correspond with states that are partly localized outside the ring. Figs. \ref{fig12}(a,b) show the asymmetry between the electron and hole states,
caused by the bias. It is seen that the electron and hole energies are related by $E_h(m,B_0) = -E_e(-m,-B_0)$, where the indices $h(e)$ refer
to holes (electrons). In the absence of bias, the electron-hole symmetry is restored, as shown in Fig. \ref{fig13}, for a ring
with $R = 50$ nm and the parabolic energy spectrum is recovered with zero energy gap.
\begin{figure}
\centering
\includegraphics[width=8 cm]{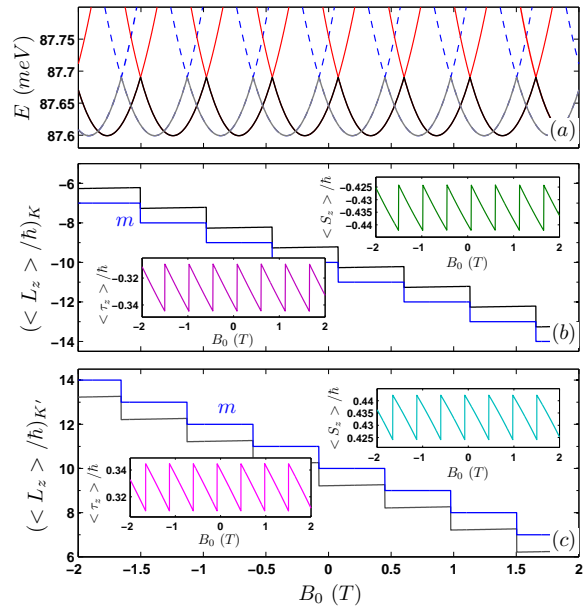}
\caption{(Color online) (a) Electron energy levels of a bilayer graphene quantum ring as function of external magnetic field $B_{0}$ for a quantum ring of radius $R=50$ nm and with $U_{b}=100~meV$ for both the $K$-valley (solid curves) and the $K'$-valley (dashed curves). Black curve shows the ground state energy of the energy spectrum in $K$-valley whereas the gray curve the corresponding ground state energy of the $K'$-valley (b) Ground state expectation values of $L_{z}/\hbar$, $S_{z}/\hbar$, $\tau_{z}/\hbar$ as function of magnetic field in the $K$-valley. Blue solid curve shows the expectation value of $J_{z}/\hbar$ operator. (c) The same as (b) but, for $K'$-valley.}
\label{fig15}
\end{figure}
\begin{figure}
\centering
\includegraphics[width=8 cm]{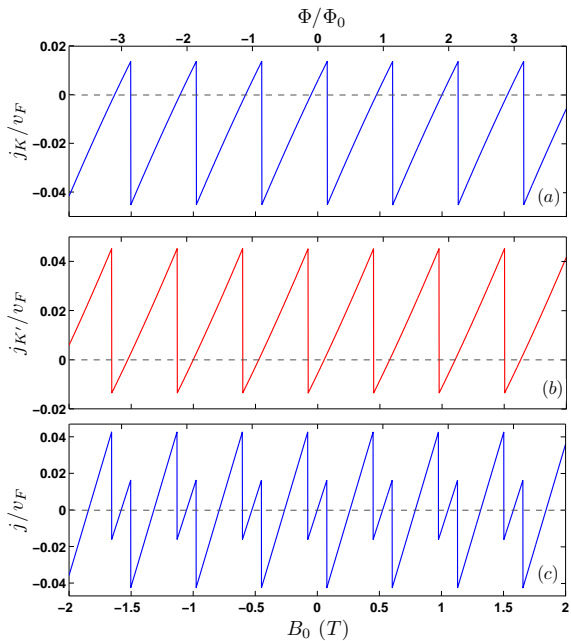}
\caption{The ground state angular current density in the (a) $K$-valley, (b) $K'$-valley and (c) the total current density of a bilayer graphene quantum ring as function of external magnetic field $B_{0}$ with $U_{b}=100$ meV and $R=50$ nm.}
\label{fig16}
\end{figure}

In Fig. \ref{fig14}, the energy branches are plotted as function of the bias, for both the
zero field case (left panels) and for $B_0 = 1$ T (right panels), with $m\geq0$ (upper panels)
and $m<0$ (lower panels). Notice that the figures are quantitatively similar to those found
previously for a quantum ring made of a single layer of graphene where the gate potential $U_{b}$
has a similar effect as the mass term $\Delta$. The differences are that for $B_{0}=0$ T the
degeneracies are now: $i$) $E(0)=E(1)=E(-1)$ and $ii$) $E(m)=E(-m)$ for $|m|>1$. In the
presence of the magnetic field  a gap is opened even for $U_b=0$ meV, which is more clearly
illustrated in the inset of the right-bottom panel of Fig. \ref{fig14}. Notice that here we
found that for $m=-1$ and $m=-2$ no real energy solution is found for $U_{b}$ below some critical value.

Figs. \ref{fig15}(b,c) shows the ground state expectation value of the angular momentum
versus the magnetic field together with the quantum number $m$ (blue solid curve) which
is an eigenvalue of the total momentum operator $J_z$. Notice that the expectation value
of $J_{z}$, i.e. $m$, is different in the $K$ and $K'$ valley which was not the case for
monolayer graphene. The energy levels for the $K$ (solid red curves) and $K'$
(dashed blue curves) valleys are depicted in Fig. \ref{fig15}(a). Black curve
(gray curve) shows the ground state energy for the $K$-valley ($K'$-valley). Notice
that in the considered case we find that the difference between $<J_z>=m\hbar$
and $<L_z>$ is about $(0.7-0.8)\hbar$ for both $K$ and $K'$ valleys.

The ground state angular current of a bilayer graphene as function of magnetic
field $B_{0}$ in the $K$-valley $j_{K}$, the $K'$-valley $j_{K'}$ and the total
angular current $j$ is shown respectively in Fig \ref{fig16}(a), (b) and (c).
In the case of a bilayer graphene quantum ring, the energy levels in the vicinity
of the $K$ and $K'$ points are different because of the valley splitting and
consequently the total angular current versus magnetic field is a more complicated
sawtooth function. Notice that the angular current for the $K$ or $K'$ valley is
not zero at $B_{0}=0$ which is due to the valley polarization whereas the total current is zero at $B_{0}=0$.

\section{Summary and conclusions}
In summary we considered the behavior of carriers in single and bilayer graphene
quantum rings within a toy model. Our approach leads to analytic expressions for the
energy spectrum. In our simple model we are not faced with
the disadvantages of the nature of edge effects which appears in quantum rings created by
cutting the layer of graphene (or lithography defined quantum rings).

We found an interesting new behavior in the presence of a perpendicular magnetic
field, which has no analogue in semiconductor-based quantum rings.
In single layer graphene quantum rings only for $\Delta>\hbar v_{F}/2R$ we found the opening of a
gap in the energy spectrum between the electron and hole states. For both single layer and
bilayer graphene quantum rings the eigenvalues are not invariant under a $B_{0}\rightarrow -B_0$ transformation  and in the case of bilayer the spectra for a fixed total angular momentum index $m$, their field dependence is  not parabolic, but exhibit two minima separated by a saddle point.
The persistent current exhibits oscillations as function of the magnetic field with period $\Phi_{0}/\pi R^2$ which are the well-known Aharonov-Bohm oscillations. Because of the valley splitting in the energy spectrum of bilayer graphene the total current density versus magnetic field is a more complicated sawtooth function.\\

{\it Acknowledgements}
This work was supported by the Flemish Science Foundation (FWO-Vl), the Belgian
Science Policy (IAP), the Bilateral program between Flanders and Brazil,
and the Brazilian Council for Research (CNPq).


\end{document}